\def\year{2019}\relax
\documentclass[letterpaper]{article} % DO NOT CHANGE THIS
\usepackage{aaai19}  % DO NOT CHANGE THIS
\usepackage{times}  % DO NOT CHANGE THIS
\usepackage{helvet} % DO NOT CHANGE THIS
\usepackage{courier}  % DO NOT CHANGE THIS
\usepackage[hyphens]{url}  % DO NOT CHANGE THIS
\usepackage{graphicx} % DO NOT CHANGE THIS
\usepackage{graphicx}  % DO NOT CHANGE THIS
\usepackage{mathtools}
\usepackage{microtype}
\usepackage{balance} 
\usepackage[export]{adjust box}
\usepackage{pgf}
\usepackage{balance} 
\usepackage{multirow}
\usepackage{hyperref}
\title{Classifying Relevant Social Media Posts During Disasters  \\ 
Using Ensemble of Domain-agnostic and Domain-specific Word Embeddings} 
\author{Ganesh Nalluru, Rahul Pandey, Hemant Purohit\\
Volgenau School of Engineering, George Mason University\\
Fairfax, VA, 22030\\
\{gn, rpandey4, hpurohit\}@gmu.edu\\
}
\begin{document}
\pubnote {\emph{AAAI 2019 Fall Symposium Series: AI for Social Good}}
\maketitle

\begin{abstract}
The use of social media as a means of communication has significantly increased over recent years. There is a plethora of information flow over the different topics of discussion, which is widespread across different domains. The ease of information sharing has increased noisy data being induced along with the relevant data stream. Finding such relevant data is important, especially when we are dealing with a time-critical domain like disasters. It is also more important to filter the relevant data in a real-time setting to timely process and leverage the information for decision support. 
% Social media has become an integral part of our communication during time-critical events. Both the affected and remotely situated citizens share a variety of posts on social media platforms with textual  content that often include information on resource needs to damage reports. Extracting such relevant posts may provide valuable situational awareness information to emergency services, however, the information overload of big social data challenges the timely processing and extraction of relevant information.  

However, the short text and sometimes ungrammatical nature of social media data challenge the extraction of contextual information cues, which could help differentiate relevant vs. non-relevant information. This paper presents a novel method to classify relevant social media posts during disaster events by ensembling the features of both domain-specific word embeddings as well as more generic domain-agnostic word embeddings. Therefore, we develop and evaluate a hybrid feature engineering framework for integrating diverse semantic representations using a combination of word embeddings to efficiently classify a relevant social media post. 
The application of the proposed classification framework could help in filtering public posts at large scale, given the growing usage of social media posts in recent years. 
\end{abstract}
\section{Introduction}\label{sec:intro}
%%%%%%%%%%%%%%%%%%%%%%%%%%%%%%%%%%%%%%%%%%%%%%%%%%%%%%%%

Social media has become one of the most effective ways of communication and information sharing during time-critical events like disasters. Due to its ease of availability, there is a vast amount of non-relevant posts such as jokes~\cite{imran2015processing} shared in the name of the disaster that makes the process of finding relevant information~\cite{purohit2013emergency} a challenging task. 
% Social media plays a key role during disaster events, as evident from the events in recent years. People share all kinds of information, ranging from jokes and prayers to damage in the affected regions~\cite{imran2015processing} to requests and offers to help~\cite{purohit2013emergency}. Studies in recent years have also shown the public expectation for timely seeking a response to their calls for help on social media~\cite{reuter2017towards}. 

The current practice across various public services for information filtering is to primarily leverage human resources through keyword-based filtering, and for emergency domains, it is done by personnel of emergency management agencies like the Public Information Officers (PIOs). Their main task is to monitor the serviceable requests for help~\cite{purohit2018social} or any actionable information coming from the public to provide relevant intelligence inputs to the decision-makers \cite{hughes2012evolving}. However, due to the burstiness of the incoming stream of social media data during the time of emergencies, it is really hard to filter relevant information given the limited number of emergency service personnel \cite{castillo2016big}. Therefore, there is a need to automatically filter out relevant posts from the pile of noisy data coming from the unconventional information channel of social media in a real-time setting. %This has been a top priority task for the emergency agencies. 
% For emergency management agencies, getting relevant situational awareness information from the affected public is of utmost importance. In particular, the Public Information Officers (PIOs) monitors the information from public to provide intelligence to the decision makers in the response coordination\cite{hughes2012evolving}. Given the increasing importance of social media data, emergency services have started to monitor social media~\cite{dhs2014using}. However, given their limited human resources and the vast amounts of social media messages posted with high velocity during disaster events, a critical challenge is to address the high information overload on the emergency service personnel~\cite{castillo2016big}.  
% There is a recognition of the necessity to effectively filter, prioritize, and organize information from this unconventional information channel~\cite{dhs2014using}. Thus, quickly filtering and prioritizing such social media posts with relevant information have become a critical need for response agencies~\cite{dhs2014using}. % 

Our contribution:
We provide a generalizable classification framework to classify relevant social media posts for emergency services.  
We introduce the framework in the 
\hyperref[sec:approach]{Method} 
section, including the description of domain-agnostic and domain-specific embeddings, learning models, and baselines. We then demonstrate the preliminary evaluation in the 
\hyperref[sec:experiments]{Experiments and Results}  
section, by experimenting with the real-world Twitter data collected during three disaster events. Lastly, we conclude with lessons and future directions in the \hyperref[sec:conclusion]{Conclusions} section. 

%%%%%%%%%%%%%%%%%%%%%%%%%%%%%%%%%%%%%%%%%%%%%%%%%%%%%%%%
\section{Related Work}\label{sec:related}
%%%%%%%%%%%%%%%%%%%%%%%%%%%%%%%%%%%%%%%%%%%%%%%%%%%%%%%%

Prior research on using social media for disaster management focuses on issues such as high volume, variety, and velocity of the incoming streams of data~\cite{castillo2016big}. Researchers have also tried to combine different modalities to filter information from the social data~\cite{nalluru2019relevancy,alam2018crisismmd}. The prior research on crisis informatics~\cite{palen2016crisis} has looked into a different perspective of computing, information, and social sciences when investigating the use of social media. 
% There has been extensive research on the topic of social media for emergency management in the last decade~\cite{imran2015processing,castillo2016big}. The nature of data generated over social media has such a high volume, variety, and velocity causing the challenges of ``Big Crisis Data'' that often overwhelm the emergency services~\cite{castillo2016big}. The literature in crisis informatics~\cite{palen2016crisis} field has investigated social media for emergency services using diverse multidisciplinary perspectives. % of computing, information, and social sciences.  
% %The emergency services have Public Information Officers (PIOs) as the critical actors who monitor social media as well as serve information to the public for helping the operational response of agencies~\cite{dhs2014using}. 
% User studies with emergency responders have identified information overload as one of the key barrier for efficiently using social media platforms by PIOs and emergency services~\cite{hiltz2014use,castillo2016big}. Such information overload factors include the processing of unstructured and noisy nature of multimodal social media content at large scale, which is beyond the capacity of the limited human resources. Furthermore, characterizing the relevancy of social media content is very contextual, time-sensitive, and often challenging~\cite{purohit2018social}.  

Understanding the text of social media has been challenging for various classification schemes. Researchers have utilized different rule-based features as well as more social media-centric features for information filtering \cite{agichtein2008finding,mostafa2013more,sriram2010short}. Moreover, there has been a rise in using word embeddings that captures more semantic relationships among texts for efficient short-text classification \cite{tang2014learning,yang2018using}. 

Word embeddings can be categorized into two types. The first category is the domain-agnostic word embeddings that are trained on a much larger domain-independent text corpus, e.g., word2vec~\cite{mikolov2013distributed}, GloVe~\cite{pennington2014glove}, and FastText~\cite{bojanowski2017enriching}. These embeddings contain broader semantic information among the words and can capture better dependencies. However, they lack the information of Out-of-Vocabulary (OOV) words and their dependencies among other words. The other category of word embeddings is the domain-specific word embeddings, which are trained on domain-specific text corpora, e.g., CrisisNLP resource~\cite{imran2016lrec} for disaster domain. This word embedding representation captures more rigorous domain-specific dependencies and information of the words. However, since the data corpus on which they are trained on was small, they sometimes lack the generic information of the words. 
% Among the social media analytics approaches, researchers have modeled public behavior in specific emergencies, addressed the problems of data collection and filtering, classification and summarization as well as visualization of analyzed data for decision support~\cite{imran2015processing}. However, the focus of such works has centered around text analytics except recent studies~\cite{alam2018twitter,alam2018crisismmd,nguyen2017damage,basnyat2017analyzing} on processing multimedia content of the social posts. Although current multimodal information processing approaches for social media mining during disasters primarily analyzed only the damage assessment aspect of emergency management.  

Based on the above studies, we propose a generic classification framework for relevant information that exploits both domain-agnostic and domain-specific word embeddings for efficient classification of social media posts through a novel ensembling approach. 

%%%%%%%%%%%%%%%%%%%%%%%%%%%%%%%%%%%%%%%%%%%%%%%%%%%%%%%%
\section{Method}\label{sec:approach}

% \subsection{Summary of the Proposed Approach}

% Citizens share a lot of information in twitter during disasters, which includes both informative and non-Informative messages. The purpose of this research is to classify informative vs Non-Informative messages and filter out the Non-Informative messages.The messages are related to crisis and it is important to use the crisis word embedding which helps in better understanding context of these messages.But,Considering only crisis word embedding does not perform better for which we propose a text representation of text that considers both crisis embedding and more generalized (glove,fast-text) embedding.Our approach is taking ensemble of a predictions from different word embeddings so that model tries to capture domain-specific context from crisis embedding and more generalized context from glove and fast-text embedding. 
In this section, we describe in detail our proposed approach. We start with the different text features we have utilized. Next, we give a detailed description of the different
learning models used. Finally, we explain our classification
framework used for ensembling the different text features for
efficient learning.

\subsection{Extracting Text Features}
In this section, we describe different methods used to derive the text features. We have used three pre-trained embeddings for this, where two embeddings (FastText and GloVe) are trained on domain-agnostic general data in the English language while one embedding (Crisis) is trained on disaster domain-specific data. % from multimodal data in the final ensemble. 

\textbf{GloVe Embedding:} GloVe or Global Vector for word Representation~\cite{pennington2014glove} learns the word embedding representation from word-word co-occurrence matrix. We have used the GloVe trained on 2 billion tweets with 27 billion tokens of 1.2 million vocabularies. 

\textbf{FastText Embedding:}
FastText~\cite{bojanowski2017enriching} uses Continuous-Bag-of-Words model with a position-weights, in dimension 300, with character n-grams of length 5, a window of size 5 and 10 negatives. FastText also represents a text by a low dimensional vector, which is obtained by summing vectors corresponding to the words appearing in the text. We have used FastText for the English language from word vectors for 157 languages\footnote{https://fasttext.cc/docs/en/crawl-vectors.html}.

\textbf{Crisis Embedding:}
In crisis embedding~\cite{imran2016lrec}, authors trained a Continuous-Bag-of-Words model with pre-trained word-embedding initialized at the beginning. They trained the model on 
a large crisis dataset from their AIDR platform~\cite{imran2014aidr}.  

\begin{figure*} 
\centering
\includegraphics[width=7in]{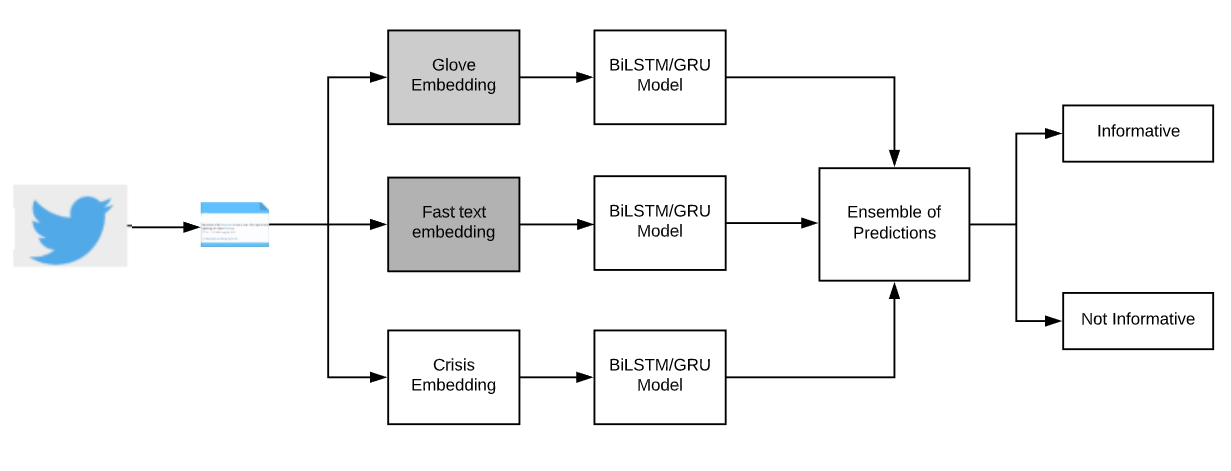}
\caption{Proposed classification framework that exploits both ensemble of word embeddings trained on LSTM/GRU model.} 
\label{fig:correlation}
\end{figure*}

\subsection{Model Building} 
Next, we discuss the deep learning sequence models used and advantages of using our proposed model. 

%Traditional neural networks do not take sequence data representation into account. This means such an approach considers all inputs and outputs as independent entities, so they are not suitable for predicting sequences. For example, if a next word has to be predicted in a sentence, at least a few words before the target word (to be predicted) should be taken into account. 

Sequence models take outputs of previous layers into account and   predictions are made based on the conditional probability of the next word given all the previous words. 

RNN is commonly used sequence model where information from previous time points are used as input for the next point, and when predicting label $y$, the network uses information not only from the corresponding input $x$ but also  from the previous layers. The drawbacks of using RNN is that it suffers from vanishing gradients problem, which cannot capture long-term dependencies, as it is difficult to update the weights of previous layers during backpropagation.

% For example, consider this tweet from the Hurricane Harvey data:\textit{``What's more IMPORTANT @realDonaldTrump? BILLIONS to FUND MEXICO BORDER WALL or REBUILDING TEXAS in WAKE of HURRICANE \_url\_ 
% % https://t.co/myHT1iOFku
% ''} and in this case RNN’s might suffer from ( short-term memory) as it is tough for them to carry information from earlier time steps to a later one.

So, to tackle the vanishing gradient problem that a normal RNN model suffers, LSTM (Long Short Term Memory) and GRU (Gated Recurrent Unit) models are used. These algorithms solve the vanishing gradient problem by implementing gates within their network, enabling them to capture much longer range dependencies and regulate the flow of information in a sequence.  

\subsubsection{BI-LSTM and GRU Models:} 
LSTM has three gates that help to control the information flow in a cell, namely Forget gate, Input gate, and Output gate. 

\textit{Forget Gate:} Forget gate helps us to decide if information should be forgotten or kept. After getting the output of previous state, % $h(t-1)$,
the current input is passed through the sigmoid function, which helps to squeeze values between 0 and 1, and based on this, the forget gate decides if information should be kept or forgotten.
% \begin{equation}
%     f_t = \sigma (W_f .[h_{t-1}, x_t] + b_f )
% \end{equation}

\textit{Input Gate:} Input gate helps us to update the cell state.  The previous hidden state and current input is passed into a sigmoid function that helps us decide which word is important and then fed into a tanh function.
% \begin{align}
% % \begin{aligned}
%     i_t = \sigma (W_i .[h_{t-1}, x_t] + b_i ) \\
%     \tilde{C}_t = tanh(W_C .[h_{t-1}, x_t] + b_C )
% % \end{aligned}
% \end{align}

\textit{Output gate:} The output gate decides the next hidden state by applying the sigmoid function and then multiplies the input with tanh to squeeze the values between -1 and 1.Finally,relevant information is sent by mutiplying it with sigmoid function.

% \begin{align}
%     o_t = \sigma (W_o .[h_{t-1}, x_t] + b_o ) \\
%     h_t = o_t*tanh(C_t)
% \end{align}

LSTM does not take into account the words that come after the current word in the sequence and rather uses only the information from previous words in the sequence. Meanwhile, Bidirectional LSTM takes information from both left-to-right and right-to-left directions, henceforth covering both information for the previous words and future words into consideration at a particular timestep. For this reason, we have used Bi-LSTM model to get more contextual information.The GRU is a simpler version of LSTM since it does not have the cell state and only has two gates, a reset gate, and an update gate. 

For training the above-selected model, first, we have used features from each embedding as baselines. Moreover, we have also utilized two different formats of combining the embeddings for better knowledge representation. For that, we have used two implementations to combine the above mentioned pre-trained word embeddings: Meta Embedding and our proposed ensemble of models with different embeddings. 

\subsubsection{Meta Embedding:} Inspired from the meta embedding research from \cite{coates-bollegala-2018-frustratingly}, we take the average of all input embedding as the final representation for each word in the input tweet messages followed by training a Bi-LSTM or GRU model.

\begin{figure} 
\centering
\includegraphics[width = \columnwidth]{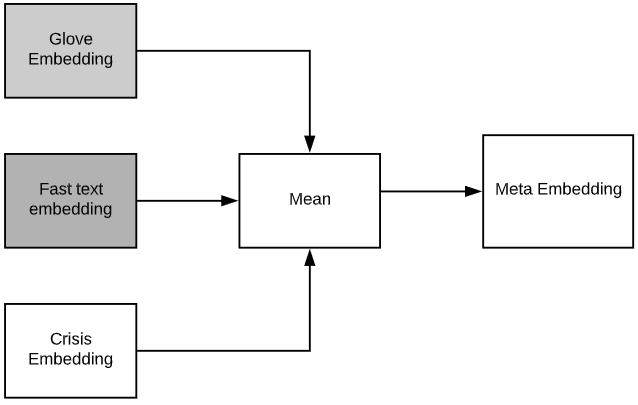}
\caption{Meta Embedding. \textit{(Average of all embeddings)} } 
\label{fig:example}
\end{figure}

\subsubsection{Ensemble of model predictions from different embeddings  (proposed):} In this method, we take the ensemble of all predictions from different models, each trained on different word embeddings, respectively. Each model is trained independently. We hypothesize that each feature set should learn on their own. Hence, for test data, if one classifier is uncertain about its predictions, the other classifiers act as a catalyst to get the correct final prediction. Each model was given different weights based on how the model performed on different word embeddings. For example,based on experiments for Hurricane Maria data with GRU
model, predictions from fasttext and Glove based GRU models were given weights of 0.4 respectively, whereas crisis embedding based GRU model was given 0.2. Hence, to calculate final predictions $P_{final}$, we will use the equation \ref{eqn:final_pred} given below. 
\begin{equation}
\label{eqn:final_pred}
P_{final} = (0.4\times P_{glove})
+ (0.2\times P_{crisis})
+ (0.4\times P_{fasttext})
\end{equation}

\section{Experiments and Results}\label{sec:experiments}
\begin{table*}
  \caption{Comparison of accuracy for various training schemes. The results from the proposed ensemble of embeddings (scheme with *) outperform other schemes across all disaster events.} 
  
  \label{tab:result} 
  \centering
  \begin{tabular}{|p{1.5cm}|p{8.0cm}|r|r|} 
    \hline  
    \multirow{1}{*}{\textbf{Dataset}} & \multirow{1}{*}{\textbf{Input Features}} & 
    {\begin{minipage}{2.5cm} \bf  [\textit{M1}] Bi-LSTM \end{minipage}} &
    {\begin{minipage}{2.0cm} \bf  [\textit{M2}] GRU \end{minipage}} \\
    % &
    % \begin{minipage}{1.3cm} \bf Sufficiently Detailed \end{minipage} \\ \hline
%   & & \bf Accuracy & \bf AUC & \bf Accuracy & \bf AUC & \bf Accuracy & \bf AUC \\ 
    \hline
   \multirow{5}{1.2cm}{{Hurricane Maria}} & {[\textit{T1}] GloVe }   &  78.64\%  & 76.67\% \\ 
   & {[\textit{T2}] FastText }   & \textbf{79.29\%}  & 75.57\% \\ 
   & {[\textit{T3}] Crisis}   &  76.34\%  & 76.01\% \\ 
   & {[\textit{M(T1+T2+T3)}] meta(GloVe + FastText + Crisis)}   
   &  78.09\%  & 76.01\% \\ 
   & {[\textit{*E(T1+T2+T3)}] ensemble(GloVe + FastText + Crisis)}   & 79.18\%  &  \textbf{77.21\% }\\ \hline
   \multirow{5}{1.2cm}{{Hurricane Harvey}} & {[\textit{T1}] GloVe }   &  85.93\%  & 84.13\% \\ 
   & {[\textit{T2}] FastText }   &  85.37\%  & 84.47\% \\ 
   & {[\textit{T3}] Crisis}   &  84.7\%  & 83.91\% \\ 
   & {[\textit{M(T1+T2+T3)}] meta(GloVe + FastText + Crisis)}   &  84.58\%  & 82.67\%\\ 
   & {[\textit{*E(T1+T2+T3)}] ensemble(GloVe + FastText + Crisis)}   &\textbf   {86.61\%}  &\textbf {84.58\%} \\ \hline
   \multirow{5}{1.2cm}{{Hurricane Irma}} & {[\textit{T1}] GloVe }   &  83.64\%  & 82.43\% \\ 
   & {[\textit{T2}] FastText }   &  83.31\%  & 82.65\% \\ 
   & {[\textit{T3}] Crisis}   &   82.76\% & 82.32\% \\ 
   & {[\textit{M(T1+T2+T3)}] meta(GloVe + FastText + Crisis)}   &  81.87\%  & 82.09\% \\ 
   & {[\textit{*E(T1+T2+T3)}] ensemble(GloVe + FastText + Crisis)}   & \textbf {83.75\%}  & \textbf{83.31\%} \\ \hline
\end{tabular}
\end{table*}
For the validation of our hypothesis, we have done an extensive evaluation of our proposed approach against multiple baselines. We will explain our experimental setup and evaluation in this section. 

\subsection{Dataset}
We used the labeled dataset for three major disaster events of 2017 -- Hurricane Maria, Hurricane Harvey, and Hurricane Irma provided by~\cite{alam2018crisismmd}. 
% This dataset provides both text and corresponding image data posted on twitter during disasters. There are both binary classification task like informative vs non-informative as well as multiclass classification tasks like humanitarian categories classification or damage severity assessment.
For our experimentation, our goal is to classify the informative tweets (relevant) vs non-informative tweets (non-relevant) using the proposed method of ensembling domain-agnostic and domain-specific embedding representations and compare its performance when only using single embedding representation. The size of the three datasets is as follows: Hurricane Maria has 2844 informative tweets and 1718 non-informative tweets, Hurricane Harvey has 3334 and 1109 tweets, and Hurricane Irma has 3564 and 957 tweets respectively. 
We conduct standard text pre-processing such as removing URLs, special characters, symbols like `RT' or `@USER' tag, and stop-words as well as lowercasing the words. 
% We conduct our experiments on three large disaster events from the year 2017 - Hurricane Maria, Hurricane Harvey, and Hurricane Irma, where the data was provided by~\cite{alam2018crisismmd}. The dataset contains Twitter posts with both text and their associated images as well as class label: \textit{informative} or \textit{non-informative}. The total \textit{informative} and \textit{non-informative} tweets across the events were 3095 and 984 for Harvey respectively, 3320 and 889 for Irma, and 2569 and 1528 for Maria respectively. %, in the context of relevancy for emergency management. 
% We performed standard text preprocessing on tweets, including removing URLs, special symbols such as `RT @user', stop words, and lowercasing. 

\subsection{Classification Schemes}
We employ various classification schemes for training with different features and learning algorithms as follows (\textit{Lx} denotes learning model type, \textit{Tx} text feature type, \textit{M()} meta embeddings, and \textit{E()} ensemble of the embeddings): 

\begin{itemize}
    \item \textbf{[\textit{T1+L1}] Text embeddings (GloVe) + Bi-LSTM Model}  This method uses GloVe embeddings as an input embedding to train a Bi-LSTM model. After the embedding layer, we have a dropout of 30\%. After that, we add a Bidirectional LSTM layer with 300 dimension output and 30\% dropout for both input space and recurrent state. Then, we have two fully-connected layers of 1024 units with 80\% dropout for generalization and a sigmoid layer of size 2 for prediction.Finally,model is fit with 100 epochs and early stopping.
    
    \item \textbf{[\textit{T1+L2}] Text embeddings (GloVe) + GRU Model} This method uses same features of T1 and train on GRU model.
    \item \textbf{[\textit{T2+L1}] Text Embeddings (FastText) + Bi-LSTM Model} This method uses FastText embeddings to generate text features (T2) and train that on Bi-LSTM.
    \item \textbf{[\textit{T2+L2}] Text Embeddings (FastText) + GRU  Model} This method uses same features of T2 and train that on GRU model. 
    \item \textbf{[\textit{T3+L1}] Text Embeddings (Crisis) + Bi-LSTM Model}  This method uses the Crisis Embedding to generate text features (T3) and train that on Bi-LSTM.
    \item \textbf{[\textit{T3+L2}] Text Embeddings (Crisis) + GRU Model} This method uses same features of T3 and train on GRU model.
    \item  \textbf{[\textit{M(T1+T2+T3) + L1}] Text Embeddings (Meta-embeddings of GloVe, FastText, and Crisis) + Bi-LSTM  Model} This method uses the meta embedding of all three embedding as text features and train that on Bi-LSTM model. 
    \item \textbf{[\textit{M(T1+T2+T3) + L2}] Text Embeddings (Meta-embeddings of GloVe, FastText, and Crisis) + GRU Model} This method uses same features of T2, and train that on GRU model.
    \item \textit{(Proposed)} \textbf{[\textit{E(\{T1+T2+T3\} + L1)}] Ensemble of all 3 classifiers of (GloVe + Bi-LSTM), (FastText + Bi-LSTM), and (Crisis + Bi-LSTM)} This method ensembles all three embeddings types and train a Bi-LSTM model.
    \item \textit{(Proposed)} \textbf{[\textit{E(\{T1+T2+T3\} + L2)}] Ensemble of all 3 classifiers of (GloVe + GRU), (FastText + GRU), and (Crisis + GRU)} This method ensembles all three embedding types and train a GRU model.
\end{itemize}

\subsection{Results}
For evaluation, we have reported an accuracy score on the 20\% test data based on the 80-20\% split strategy. Table~\ref{tab:result} shows the results of our proposed method with the combinations of an ensemble of all embeddings and their comparison to the other baselines. 

As seen in the table, our proposed approach have given a better score than the baselines in all the datasets. Our final proposed approach uses an ensemble of both domain-specific (crisis) embedding and more generalized (glove, fast-text) embedding; each being trained with Bidirectional LSTM and GRU network model. Hence, it is better able to capture the domain-specific as well as generic dependencies among the textual words for each class of data, i.e. \textit{informative} vs \textit{not-informative}. We can also observe that training a different model for each embedding is better than combining all embeddings first and then training a single model (meta embedding). The reason for that can be the difference in the feature space in which each embedding has been trained on. Hence, averaging the features may lose some information or add noise what one dimension is representing in each feature embedding.  
% We have built several deep learning models based on our ensemble feature representation (combining glove, fasttext and crisis) and compared against the baseline of single embedding only features and meta-embedding. 
We can also see that all experiments with Bi-LSTM model are slightly better than their corresponding GRU model.

Furthermore, as we can infer from the results table, accuracy of models with single pre-trained embedding are almost the same, and a model's accuracy improves when we take the ensemble of different model predictions.  
% For example ,Let us consider Hurricane Irma with GRU model,We can see that model accuracy of GRU with glove embedding is 81.10\%,FastText embedding is 83.09\% and crisis embedding is 82.71\% and model improved by an accuracy of 1.5\% with ensembling.

Also, the model with only domain-specific crisis embeddings perform worse on most of the cases when compared to Glove and FastText embeddings; but, it gives a boost to the other embeddings when combined. Thus, this finding suggests the usefulness of combining both domain-specific and domain-agnostic  representations to provide more context to the learning algorithm.

\section{Conclusion and Future Work}\label{sec:conclusion}
%%%%%%%%%%%%%%%%%%%%%%%%%%%%%%%%%%%%%%%%%%%%%%%%%%%%%%%%

This paper presented a novel approach to classify relevant social media posts by ensembling domain-agnostic and domain-specific word embedding representations, which can capture different types of information in a given text and provide more context to the learning model. We have validated our approach with three disaster event datasets and achieved over 86\% accuracy score. We have also observed better performance on Bi-LSTM based model compared to GRU based model for our use case. In light of this preliminary investigation, the application of the proposed method has the potential to help improve social media services at emergency response organizations. 

In future work, we can generalize our ensemble approach more according to the training data. More specifically, since we have fixed the ratio of the prediction score coming from different embedding-based learning at the beginning in our proposed approach, we can also learn this ratio as a part of the training. We can also add more embeddings for enhancing the contextual representation of more information in the text. 

%%%%%%%%%%%%%%%%%%%%%%%%%%%%%%%%%%%%%%%%%%%%%%%%%%%%%%%%

\section{Acknowledgment}
Authors thank US National Science Foundation (NSF) grants IIS-1657379 and IIS-1815459 for partial support. 

\balance{} 
\bibliographystyle{aaai.bst}
\bibliography{fss-meta-embedding}

\end{document}